\documentclass[floats,floatfix,showpacs,amssymb,prd,twocolumn,superscriptaddress,nofootinbib,nolongbibliography,reprint,prd,aps]{revtex4-2}
\usepackage{textcomp}
\usepackage{amssymb,amsmath,verbatim,mathtools,needspace,enumitem,etoolbox,graphicx,microtype,afterpage,bigints,gensymb,tabularx,xspace,siunitx}
\usepackage{aasmacros}
\usepackage[commandnameprefix=ifneeded, authormarkup=none,final]{changes}

\definechangesauthor[name={AU}, color=red]{au} 
\newcommand{\aur}[2]{\replaced[id=au]{#1}{#2}} 
\newcommand{\aua}[1]{\added[id=au]{#1}} 

\usepackage{graphicx}
\usepackage{dcolumn}
\usepackage{bm}
\usepackage{lipsum}
\usepackage{physics}
\usepackage{soul}
\usepackage{array}
\usepackage{makecell}
\usepackage{tabularx}
\usepackage{multirow}
\usepackage{placeins}
\usepackage[normalem]{ulem}

\AtBeginDocument{\RenewCommandCopy\qty\SI}
\ExplSyntaxOn
\msg_redirect_name:nnn { siunitx } { physics-pkg } { none }
\ExplSyntaxOff
\DeclareSIUnit\solarmass{\ensuremath{M_{\odot}}}
\DeclareSIUnit\parsec{pc}
\DeclareSIUnit\lightspeed{$c$}
\DeclareSIUnit\year{yr}
\DeclareSIUnit\arcsecond{as}
\DeclareSIUnit\astronomicalunit{AU}
\DeclareSIUnit\clight{\ensuremath c}

\definecolor{linkcolor}{rgb}{0.0,0.3,0.5}
\usepackage[colorlinks = true,
            linkcolor = linkcolor,
            urlcolor  = linkcolor,
            citecolor = linkcolor,
            anchorcolor = linkcolor]{hyperref}
\usepackage[capitalize]{cleveref}
\Crefname{equation}{Eq.}{Eqs.}
\Crefname{figure}{Fig.}{Figs.}
\Crefname{tabular}{Tab.}{Tabs.}
\Crefname{section}{Sec.}{Secs.}
\Crefname{subsection}{Sec.}{Secs.}

\usepackage{array}

\usepackage{orcidlink}
\begin{document}
\sisetup{range-phrase=-, range-units=single}

\title{Accurate and efficient simulation-based inference\\ for massive black-hole binaries with LISA}

\newcommand{\milan}{\affiliation{Dipartimento di Fisica ``G. Occhialini'', Universit\'a degli Studi di Milano-Bicocca, Piazza della Scienza 3, 20126 Milano, Italy}}
\newcommand{\infn}{\affiliation{INFN, Sezione di Milano-Bicocca, Piazza della Scienza 3, 20126 Milano, Italy}} 
\newcommand{\bham}{\affiliation{Institute for Gravitational Wave Astronomy \& School of Physics and Astronomy, University of Birmingham, Birmingham, B15 2TT, UK}}

\author{Alice Spadaro~\orcidlink{0009-0000-8104-6171}}
\email{a.spadaro3@campus.unimib.it}
\milan \infn

\author{Jonathan Gair~\orcidlink{0000-0002-1671-3668}}
\affiliation{Max Planck Institute for Gravitational Physics (Albert Einstein Institute), Am Mühlenberg 1, 14476 Potsdam, Germany}

\author{Davide Gerosa~\orcidlink{0000-0002-0933-3579}}
\milan \infn

\author{Stephen R. Green~\orcidlink{0000-0002-6987-6313}}
\affiliation{Nottingham Centre of Gravity \& School of Mathematical Sciences, University of Nottingham, University Park, Nottingham, NG7 2RD, United Kingdom}

\author{\\Riccardo Buscicchio~\orcidlink{0000-0002-7387-6754}}
\milan \infn\bham

\author{Nihar Gupte~\orcidlink{0000-0002-7287-5151}}
\affiliation{Max Planck Institute for Gravitational Physics (Albert Einstein Institute), Am Mühlenberg 1, 14476 Potsdam, Germany}

\author{Rodrigo Tenorio~\orcidlink{0000-0002-3582-2587}}
\milan \infn

\author{Samuel Clyne~\orcidlink{0009-0007-2357-9958}}
\affiliation{Department of Physics and Center for Computational Research, East Hall, University of Rhode Island, Kingston, RI 02881}

\author{\\Michael Pürrer~\orcidlink{0000-0002-3329-9788}}
\affiliation{Department of Physics and Institute for AI \& Computational Research, East Hall, University of Rhode Island, Kingston, RI 02881}

\author{Natalia Korsakova~\orcidlink{0000-0002-1112-8830}}
\affiliation{Universit\'e C\^ote d’Azur, Observatoire de la C\^ote d’Azur, CNRS, Artemis, Bd de l’Observatoire, F-06304 Nice, France}

\date{\today}

\newcommand{\LowInjMc}{\ensuremath{1.61}}
\newcommand{\LowInjq}{\ensuremath{0.82}}
\newcommand{\LowInjChiOne}{\ensuremath{-0.47}}
\newcommand{\LowInjChiTwo}{\ensuremath{-0.11}}
\newcommand{\LowInjdL}{\ensuremath{44.14}}
\newcommand{\LowInjThetaJN}{\ensuremath{1.77}}
\newcommand{\LowInjThetaS}{\ensuremath{0.14}}
\newcommand{\LowInjPhiS}{\ensuremath{3.09}}
\newcommand{\LowInjTc}{\ensuremath{820.0}}
\newcommand{\LowInjPsi}{\ensuremath{1.13}}
\newcommand{\LowInjPhiC}{\ensuremath{3.36}}

\newcommand{\LowJSDMc}{\ensuremath{4.8\times10^{-4}}}
\newcommand{\LowJSDq}{\ensuremath{7.9\times10^{-4}}}
\newcommand{\LowJSDChiOne}{\ensuremath{3.7\times10^{-4}}}
\newcommand{\LowJSDChiTwo}{\ensuremath{2.4\times10^{-4}}}
\newcommand{\LowJSDdL}{\ensuremath{1.4\times10^{-3}}}
\newcommand{\LowJSDThetaJN}{\ensuremath{4.1\times10^{-3}}}
\newcommand{\LowJSDThetaS}{\ensuremath{2.3\times10^{-3}}}
\newcommand{\LowJSDPhiS}{\ensuremath{3.2\times10^{-3}}}
\newcommand{\LowJSDTc}{\ensuremath{5.1\times10^{-4}}}
\newcommand{\LowJSDPsi}{\ensuremath{1.8\times10^{-3}}}
\newcommand{\LowJSDPhiC}{\ensuremath{1.3\times10^{-4}}}

\newcommand{\LowJSDthrMc}{\ensuremath{1.0\times10^{-3}}}
\newcommand{\LowJSDthrq}{\ensuremath{1.2\times10^{-3}}}
\newcommand{\LowJSDthrChiOne}{\ensuremath{8.0\times10^{-4}}}
\newcommand{\LowJSDthrChiTwo}{\ensuremath{6.0\times10^{-4}}}
\newcommand{\LowJSDthrdL}{\ensuremath{3.0\times10^{-3}}}
\newcommand{\LowJSDthrThetaJN}{\ensuremath{7.0\times10^{-3}}}
\newcommand{\LowJSDthrThetaS}{\ensuremath{4.0\times10^{-3}}}
\newcommand{\LowJSDthrPhiS}{\ensuremath{5.0\times10^{-3}}}
\newcommand{\LowJSDthrTc}{\ensuremath{1.0\times10^{-3}}}
\newcommand{\LowJSDthrPsi}{\ensuremath{3.0\times10^{-3}}}
\newcommand{\LowJSDthrPhiC}{\ensuremath{5.0\times10^{-4}}}

\newcommand{\ModInjMc}{\ensuremath{1.30}}
\newcommand{\ModInjq}{\ensuremath{0.54}}
\newcommand{\ModInjChiOne}{\ensuremath{-0.09}}
\newcommand{\ModInjChiTwo}{\ensuremath{0.66}}
\newcommand{\ModInjdL}{\ensuremath{21.91}}
\newcommand{\ModInjThetaJN}{\ensuremath{2.33}}
\newcommand{\ModInjThetaS}{\ensuremath{1.62}}
\newcommand{\ModInjPhiS}{\ensuremath{4.72}}
\newcommand{\ModInjTc}{\ensuremath{395.0}}
\newcommand{\ModInjPsi}{\ensuremath{0.98}}
\newcommand{\ModInjPhiC}{\ensuremath{2.18}}

\newcommand{\ModJSDMc}{\ensuremath{1.2\times10^{-3}}}
\newcommand{\ModJSDq}{\ensuremath{1.6\times10^{-3}}}
\newcommand{\ModJSDChiOne}{\ensuremath{9.5\times10^{-4}}}
\newcommand{\ModJSDChiTwo}{\ensuremath{8.1\times10^{-4}}}
\newcommand{\ModJSDdL}{\ensuremath{2.9\times10^{-3}}}
\newcommand{\ModJSDThetaJN}{\ensuremath{6.4\times10^{-3}}}
\newcommand{\ModJSDThetaS}{\ensuremath{3.8\times10^{-3}}}
\newcommand{\ModJSDPhiS}{\ensuremath{4.6\times10^{-3}}}
\newcommand{\ModJSDTc}{\ensuremath{8.9\times10^{-4}}}
\newcommand{\ModJSDPsi}{\ensuremath{2.5\times10^{-3}}}
\newcommand{\ModJSDPhiC}{\ensuremath{4.0\times10^{-4}}}

\newcommand{\ModJSDthrMc}{\ensuremath{2.0\times10^{-3}}}
\newcommand{\ModJSDthrq}{\ensuremath{2.5\times10^{-3}}}
\newcommand{\ModJSDthrChiOne}{\ensuremath{1.5\times10^{-3}}}
\newcommand{\ModJSDthrChiTwo}{\ensuremath{1.3\times10^{-3}}}
\newcommand{\ModJSDthrdL}{\ensuremath{4.5\times10^{-3}}}
\newcommand{\ModJSDthrThetaJN}{\ensuremath{9.0\times10^{-3}}}
\newcommand{\ModJSDthrThetaS}{\ensuremath{6.0\times10^{-3}}}
\newcommand{\ModJSDthrPhiS}{\ensuremath{7.0\times10^{-3}}}
\newcommand{\ModJSDthrTc}{\ensuremath{1.5\times10^{-3}}}
\newcommand{\ModJSDthrPsi}{\ensuremath{4.0\times10^{-3}}}
\newcommand{\ModJSDthrPhiC}{\ensuremath{8.0\times10^{-4}}}

\newcommand{\HighInjMc}{\ensuremath{0.90}}
\newcommand{\HighInjq}{\ensuremath{0.47}}
\newcommand{\HighInjChiOne}{\ensuremath{0.34}}
\newcommand{\HighInjChiTwo}{\ensuremath{0.28}}
\newcommand{\HighInjdL}{\ensuremath{16.96}}
\newcommand{\HighInjThetaJN}{\ensuremath{0.73}}
\newcommand{\HighInjThetaS}{\ensuremath{2.94}}
\newcommand{\HighInjPhiS}{\ensuremath{3.09}}
\newcommand{\HighInjTc}{\ensuremath{-568.0}}
\newcommand{\HighInjPsi}{\ensuremath{2.51}}
\newcommand{\HighInjPhiC}{\ensuremath{2.74}}

\newcommand{\HighJSDMc}{\ensuremath{2.1\times10^{-4}}}
\newcommand{\HighJSDq}{\ensuremath{3.5\times10^{-4}}}
\newcommand{\HighJSDChiOne}{\ensuremath{1.9\times10^{-4}}}
\newcommand{\HighJSDChiTwo}{\ensuremath{1.6\times10^{-4}}}
\newcommand{\HighJSDdL}{\ensuremath{6.2\times10^{-4}}}
\newcommand{\HighJSDThetaJN}{\ensuremath{1.5\times10^{-3}}}
\newcommand{\HighJSDThetaS}{\ensuremath{9.8\times10^{-4}}}
\newcommand{\HighJSDPhiS}{\ensuremath{1.2\times10^{-3}}}
\newcommand{\HighJSDTc}{\ensuremath{2.6\times10^{-4}}}
\newcommand{\HighJSDPsi}{\ensuremath{7.1\times10^{-4}}}
\newcommand{\HighJSDPhiC}{\ensuremath{9.5\times10^{-5}}}

\newcommand{\HighJSDthrMc}{\ensuremath{5.0\times10^{-4}}}
\newcommand{\HighJSDthrq}{\ensuremath{7.0\times10^{-4}}}
\newcommand{\HighJSDthrChiOne}{\ensuremath{4.0\times10^{-4}}}
\newcommand{\HighJSDthrChiTwo}{\ensuremath{3.5\times10^{-4}}}
\newcommand{\HighJSDthrdL}{\ensuremath{1.2\times10^{-3}}}
\newcommand{\HighJSDthrThetaJN}{\ensuremath{2.5\times10^{-3}}}
\newcommand{\HighJSDthrThetaS}{\ensuremath{1.8\times10^{-3}}}
\newcommand{\HighJSDthrPhiS}{\ensuremath{2.0\times10^{-3}}}
\newcommand{\HighJSDthrTc}{\ensuremath{5.0\times10^{-4}}}
\newcommand{\HighJSDthrPsi}{\ensuremath{1.2\times10^{-3}}}
\newcommand{\HighJSDthrPhiC}{\ensuremath{2.0\times10^{-4}}}


\begin{abstract}
We develop an accurate simulation-based inference framework for high-mass ($\gtrsim\!10^7 \rm{M_\odot}$) black-hole binaries observable by LISA.
The method is implemented within the \textsc{Dingo} gravitational-wave parameter-estimation code, extending its application from ground-based detectors to the LISA band. 
We train a normalizing-flow model using aligned-spin higher-mode waveform models and a low-frequency approximation of the detector response \aua{at fixed reference time}. After sampling, we importance-sample to the true posterior \aua{based on the underlying likelihood and prior}. We validate performance on simulated signals spanning the signal-to-noise regimes relevant for LISA observations and benchmark our new \textsc{Dingo} implementation against standard methods. We report robust agreement in the inferred posterior distributions up to signal-to-noise ratios of $\sim\!500$. At higher signal-to-noise ratios of $\sim\!1000$, we observe a reduction in sampling efficiency, while still yielding unbiased and tightly localized posteriors that can be used as a starting point for follow-up with traditional methods.
The trained flow can generate 20 thousand posterior samples in less than a minute, establishing \textsc{Dingo} as a promising neural inference framework for rapid full-parameter estimation of massive black-hole binaries in the LISA band.
\aur{The proposed}{The likelihood-free nature of this} approach allows for straightforward generalizations, including a time-dependent detector response, non-stationary noise artifacts such as gaps and glitches, and low-latency parameter estimations.

\end{abstract}

\maketitle
\section{\label{sec:a} Introduction}
The Laser Interferometer Space Antenna (LISA) will soon probe the millihertz band of the gravitational-wave (GW) spectrum.
A rich variety of astrophysical sources is expected to populate LISA’s sensitivity band, with massive black-hole (BH) binaries being one of the mission's primary scientific targets~\cite{2024arXiv240207571C}.

The analysis of massive BH binary signals relies on two complementary strategies: low-latency and global-fit pipelines.
Low-latency algorithms are designed to rapidly identify and characterize GW signals, enabling early alerts for electromagnetic observatories and supporting coordinated multi-messenger follow-up campaigns~\cite{2008ApJ...684..870K,2022PhRvD.106j3017M,2023MNRAS.521.2577P}.
Significant effort has been devoted to the development of such methods, employing both approximate and computationally efficient matched-filtering techniques~\cite{2022PhRvD.105d4007C,2022PhRvD.105d4055K,2025PhRvD.111f4079J,2024CQGra..41b5006W,2024CQGra..41x5012H,2025PhRvD.111d2009S,2025PhRvD.111j4044T,2025PhRvD.112d3010D} and approaches based on machine learning~\cite{2024PhRvD.109l3031R,2024PhRvD.110f2003H,2025CQGra..42v5001I}.
On the other hand, global-fit pipelines aim to perform a simultaneous inference of the tens of thousands of resolvable sources overlapping in the LISA data stream, jointly with the instrumental and astrophysical noise~\cite{2005PhRvD..72d3005C}.
Tackling this inherently high-dimensional inference problem is paramount for building accurate and reliable source catalogs for astrophysical, cosmological, and fundamental-physics analyses, as well as refining early warnings for follow-up observations.
Conventional Bayesian inference techniques have been successfully applied to the analysis of massive BH binary signals~\cite{2021PhRvD.103h3011M, 2021PhRvD.104d4035D,2023PhRvD.108l4045P,2023PhRvD.107l3026P,2023PhRvD.108l3029S,2024PhRvD.109j4019T,2024MNRAS.535.3283G,2025CQGra..42f5018C,2025PhRvD.112f3041M,2025PhRvD.111l4053B,2025PhRvD.112l4044P}, including global-fit approaches~\cite{2023PhRvD.107f3004L,2024PhRvD.110b4005S,2025PhRvD.111j3014D,2025PhRvD.111b4060K} tested on the ``Sangria'' LISA Data Challenge~\cite{le_jeune_2022_7132178}.
To enable large-scale inference of multi-source datasets, these approaches typically rely on waveform models that compromise physical content and computational tractability.  

Fast and accurate parameter estimation is therefore crucial for maximizing the scientific return of LISA observations of massive BH binaries.
\aua{To address the computational cost of Bayesian inference, a variety of acceleration strategies have been developed, including reduced-order and surrogate modeling methods~\cite{2015PhRvL.114g1104C, 2016PhRvD..94d4031S, PhysRevD.104.063031, 2023PhRvD.108l3025M,2026arXiv260109819N, 2022LRR....25....2T}, heterodyned likelihoods~\cite{2010arXiv1007.4820C, 2018arXiv180608792Z, 2021PhRvD.104j4054C, 2026arXiv260111239S}, and adaptive sampling techniques such as multibanding~\cite{2017CQGra..34k5006V,2021PhRvD.104d4062M}.}
Recently, active-learning approaches based on Gaussian processes have been proposed to accelerate likelihood-based inference by significantly reducing the number of costly likelihood evaluations~\cite{2025PhRvD.112f3010E}. 
Irrespective of the speed-up achievable in classical methods, it is crucial to note that the simplifying assumptions—i.e. Gaussianity and stationarity—at the core of likelihood-based inference approaches are violated in real GW data. 
Indeed, instrumental artifacts, including glitches and data gaps, further complicate the likelihood modeling, potentially introducing systematic biases if neglected~\cite{2023PhRvD.108l3029S,2024PhRvD.109h3027H,2025PhRvD.112f3041M} or significantly increasing computational costs when explicitly accounted for~\cite{2019PhRvD.100b2003B,2025PhRvD.111l4053B}.

In this broader context, simulation-based inference (SBI) has emerged as a promising likelihood-free framework for GW data analysis~\cite{2020PhRvD.102j4057G,2020arXiv200803312G}, also in the regime of space-based detectors~\cite{2024PhRvD.109h3008A,2025PhRvD.111j2006A,2025arXiv250516795C,2025JCAP...04..022M,2024MLS&T...5d5040L}.
Specifically, recent studies have applied SBI techniques to massive BH binaries, including neural likelihood estimation~\cite{2025JCAP...04..022M} designed for inference on a single observation, and neural posterior estimation~\cite{2024MLS&T...5d5040L}, which enables parameter estimation across different events.
While these works show encouraging results, their comparison with standard Bayesian techniques merits further investigation and is discussed in Sec.~\ref{previous}.

Among SBI-based approaches, the \textsc{Dingo} software has been extensively validated in GW astronomy.
Most notably, Ref.~\cite{2021PhRvL.127x1103D} demonstrated fast and accurate parameter inference for compact binary signals observed by ground-based detectors.
Subsequent works have extended the framework to neutron-star binaries~\cite{2025Natur.639...49D}, explored its performance for third-generation detectors~\cite{2025PhRvD.112j3015S}, and more recently generalized it through transformer-based architectures~\cite{2025arXiv251202968K}.
In this work, we adapt \textsc{Dingo} to perform neural posterior estimation on quasi-circular, non-precessing massive BH binaries observed by LISA. 
The signal model is based on the \textsc{IMRPhenomXHM} waveform approximant~\cite{2020PhRvD.102f4002G}, with the detector response included within the low-frequency approximation \aua{and evaluated at fixed reference time}.
A conditional normalizing-flow architecture composed of spline coupling layers~\cite{2019arXiv190604032D} is trained to approximate the target posterior distribution. 
The performance of the method is assessed via importance sampling and benchmarked against traditional stochastic samplers.

This paper is organized as follows. 
In Sec.~\ref{methods}, we present the LISA targeted sources, describe the adopted detector response, and detail the neural posterior estimation methodology used in this analysis.
In Sec.~\ref{results}, we present and discuss our results, and briefly compare them with previous work.
In Sec.~\ref{appl and challenges}, we examine the implications of our findings and outline the main challenges and limitations of the proposed approach. 

The LISA version of the \textsc{Dingo} code is under active development. 
Progress can be tracked at a dedicated fork of the main {\sc Dingo} repository~\cite{Dingo}.

\section{\label{sec:b} Methods}
\label{methods}

Our analysis targets the most massive BH binary systems observable by LISA, with redshifted total masses $M\gtrsim10^7 \rm{M_\odot}$ and signal-to-noise-ratios (SNRs) ranging from $\sim 10$ to $\sim 1000$.
Owing to the mission low-frequency cut-off at $f_{\rm{low}}=0.1~\rm{mHz}$~\cite{2024arXiv240207571C}, these binaries sweep through the LISA band over relatively short timescales, $\tau\sim\mathcal{O}(\rm{hours-days)}$.
As a consequence, LISA is expected to observe only the late inspiral, merger, and ringdown, 
spanning the final few to few tens of GW cycles, $N_{\rm GW}\sim\tau f_{\rm{low}}$.
The resulting signals exhibit a morphology similar to that of the transient events detected by the current ground-based observatories LIGO and Virgo. 
As such, the analysis of massive BH binaries observed by LISA can be naturally framed within the existing \textsc{Dingo} framework.
In the following, we describe the LISA response model implemented within \textsc{Dingo} (Sec.~\ref{lisa conventions}) and outline the corresponding training setup (Sec.~\ref{dingo-lisa}). 

\subsection{LISA response}
\label{lisa conventions}
The observables of space-based GW interferometers are obtained through time-delay interferometry (TDI), which is a post-processing technique that reduce the overwhelming laser frequency noise in the phase measurements~\cite{2021LRR....24....1T}.
In this work, we model the LISA response within the low-frequency approximation~\cite{1998PhRvD..57.7089C}, which provides a simplified yet well-motivated description for the high-mass binary signals produced by the sources considered here.
Accordingly, LISA is modeled as a rigid triangular constellation with equal arm lengths, neglecting both the time and frequency dependence of the response~\cite{2003PhRvD..67b9905C}.
Under these assumptions, the constellation is effectively treated as fixed along its orbit and the TDI variables are approximated by two noise orthogonal data channels (commonly denoted A and E), whose GW response is equivalent to that of two co-located, motionless Michelson interferometers rotated with respect to each other by an angle of~$\pi/4$. 

We assume the noise in the two channels to be identical, stationary, and Gaussian, with statistical properties fully described by a power spectral density (PSD) $S_n(f)$.
\aur{We adopt a fixed PSD based on}{We adopt} the SciRDv1 noise model~\cite{SciRD} simulated using \texttt{lisatools}~\cite{michael_katz_2024_10930980}, while neglecting the confusion noise arising from the superposition of unresolved Galactic binaries (see e.g. Ref.~\cite{2025arXiv251103604B} and references therein).
The inclusion of this additional noise component, characterized by non-Gaussian and non-stationary features, \aua{as well as a modeling of noise artifacts,} is a natural direction for future SBI studies.
The instrument response is implemented within the \textsc{Dingo} framework 
in close correspondence with the formalism of Ref.~\cite{1998PhRvD..57.7089C}.

\subsection{{\sc Dingo} setup}
\label{dingo-lisa}
We perform Bayesian neural posterior estimation in the frequency domain between $f_{\rm min} = 0.1\,\rm{mHz}$ and $f_{\rm max} = 4\,\mathrm{mHz}$, with a resolution of $\Delta f = 10^{-5}\,\rm{Hz}$.
As shown in Ref.~\cite{2021PhRvD.103h3011M}, the frequency dependence of the response vanishes in the limit 
$f\ll f_L\sim0.19\,\mathrm{Hz}$, where $f_L=c/(2\pi L)$ is the detector transfer frequency and $L$ the LISA arm length.
Although departures from the low-frequency approximation become noticeable around $2\,\mathrm{mHz}$, most of the signal content for the systems considered here lies at lower frequencies \aua{($f < 1\,\rm{mHz}$)}, so that the approximation is expected to remain accurate over the frequency range analyzed.
\aua{A detailed assessment of potential systematic effects induced by the response approximation is beyond the scope of this work and is left to future analyses.}

We perform inference over the full 11-dimensional parameter space of quasi-circular spin-aligned compact binaries
\[\theta = \{\mathcal{M}_c, q, \chi_{1,2}, d_L, \theta_{JN}, \theta_S, \phi_S, \psi, t_c, \phi_c\},\] where $\mathcal{M}_c$ is the redshifted chirp mass, $q$ is the mass ratio, $\chi_{1,2}$ are the dimensionless spin components aligned with the orbital angular momentum, 
$d_L$ is the luminosity distance, $\theta_{JN}$ is the  inclination angle between the total angular momentum and the line of sight, $\theta_S$ is the ecliptic colatitude, $\phi_S$ is the ecliptic longitude, $\psi$ is the polarization angle, $t_c$ is the coalescence time, and $\phi_c$ is the coalescence phase.
All extrinsic parameters are defined in the frame of the Solar System barycenter.

For density estimation, we employ a conditional normalizing flow based on a rational-quadratic spline coupling transforms~\cite{2019arXiv190604032D}, adopting the same network architecture as in Ref.~\cite{2021PhRvL.127x1103D}. 
In addition to the flow network, \textsc{Dingo} incorporates an embedding network that compresses the input strain data into a 128-dimensional feature vector for each detector channel.
Together, these two networks comprise approximately 361 millions learnable parameters.

We train the neural network on simulated data generated with the \textsc{IMRPhenomXHM} waveform approximant~\cite{2020PhRvD.102f4002G}, including all available emission modes and
additive stationary Gaussian noise realizations sampled from the fixed PSD. 
The training dataset includes $10^7$ waveforms covering the intrinsic-parameter space, while the extrinsic parameters are generated on-the-fly.
Parameters $\theta$ are drawn as follows: chirp mass $\mathcal{M}_c\in[8\times10^6,\,2\times10^7]\,\rm{M_\odot}$ and mass ratio $q \in [0.2,1]$ are derived from components masses sampled uniformly in the range $m_1, m_2 \in [7\times10^6,\,3\times10^7]\,\rm{M_\odot}$, subject to the constraint $m_1 > m_2$. 
This choice ensures that the signal remains within the validity range of the low-frequency response model discussed above.
Following Ref.~\cite{2020arXiv200803312G}, we adopt a uniform prior in luminosity distance, $d_L \in [15,50]\,\mathrm{Gpc}$, which improves training performance, while ensuring coverage of the target SNR range across the mass interval considered.
\begin{figure}
    \centering
    \includegraphics[width=1.\columnwidth, keepaspectratio]{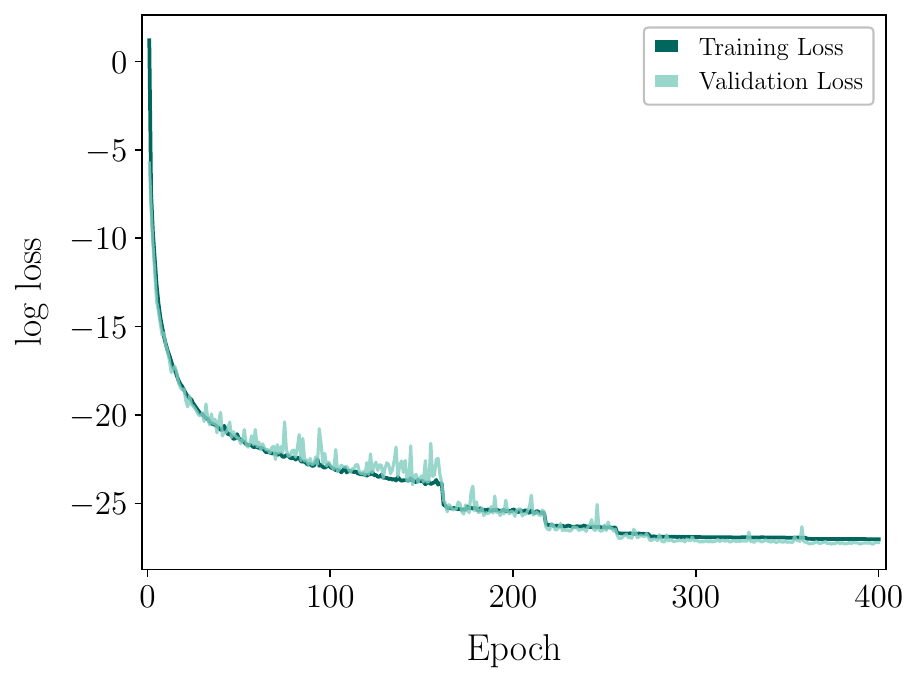}
    \caption{Evolution of the training (green) and validation (teal) losses as a function of epoch. 
    The drops at epochs $\sim$160, 220, 255, 275, and 360 are due to learning rate reductions triggered by the \texttt{ReduceLROnPlateau} scheduler~\cite{2019arXiv191201703P}.
    }
    \label{fig:loss}
\end{figure}
Spin magnitudes are taken to be uniformly distributed in $[0,0.99]$, while spin orientations are assumed to be isotropic with the planar components marginalized over.
The resulting prior on the aligned dimensionless spins $\chi_{1,2}\in [-1,1]$ peaks at zero, as shown in Fig.~\ref{fig:prior_coverage}.
Standard uninformative priors are adopted for all remaining angular parameters.
All training signals are evaluated at a fixed reference time $t_{\rm ref}=1\,\mathrm{yr}$, so that the neural network does not need to learn the modulation induced by LISA’s heliocentric motion.
As discussed in Ref.~\cite{2024MLS&T...5d5040L}, correcting for detector motion would be straightforward.
Finally, we adopt a uniform prior on the coalescence time $t_c$ within 30 minutes around the reference time $t_{\rm ref}$, implicitly assuming that a sufficiently accurate trigger time is supplied by a preceding low-latency search pipeline. 
The short runtime of our method (see Sec.~\ref{bench} for details) allows continuous processing of consecutive data segments as an alternative to a trigger-based approach, provided the model remains uninformative in the absence of a signal.

The network is trained for 400 epochs with a batch size of 4096. 
We use the Adam optimizer~\cite{kingma2017adammethodstochasticoptimization} with an initial learning rate of $2\times10^{-4}$, which is reduced by a factor of 0.4 when the validation loss stagnates for 15 epochs.
As shown in Fig.~\ref{fig:loss}, the training and validation losses closely track each other throughout optimization, indicating stable convergence and no evidence of overfitting.
Training was performed on a single 40 GB NVIDIA A100 GPU and required approximately 10 days.

\section{Results}
\label{results}

We validate our inference code both internally, using a large number of injections drawn from the prior distribution (Sec.~\ref{model validation}), and externally, by comparison with a standard stochastic sampler (Sec.~\ref{bench}) and previous works (Sec.~\ref{previous}). 

\subsection{Model validation} 
\label{model validation}
Once trained, we employ the flow-based model for posterior inference, enabling rapid sampling for data consistent with the training priors and the assumed PSD of the noise.
Residual discrepancies in the learned posterior approximation—arising from limited training or finite network expressivity—are mitigated via importance sampling (IS)~\cite{2023PhRvL.130q1403D} using the flow posterior as a proposal distribution.
We quantify the quality of the reweighted samples through the sampling efficiency
\begin{figure}[b]
    \centering
    \includegraphics[width=1.\columnwidth, keepaspectratio]{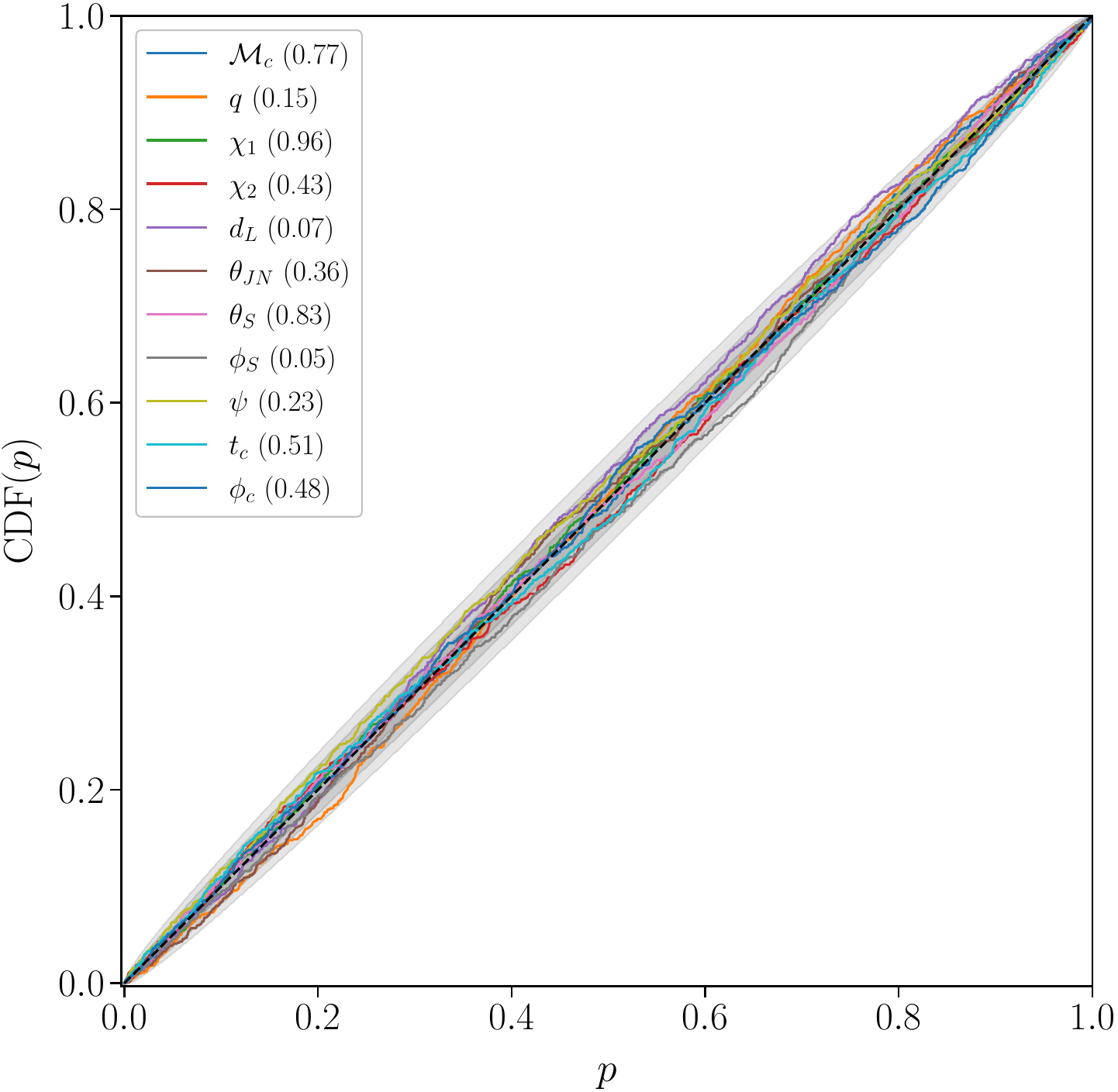}
    \caption{Probability-probability ($p$--$p$) plot for 1000 simulated injections obtained with \textsc{Dingo-IS}. Colored curves show results for each of the  11 marginal distributions. The dashed line indicates a uniform distribution, and grey shaded regions denote the expected $1\sigma,\,2\sigma,\,3\sigma$ confidence intervals. KS test $p$-values are provided in the legend. 
    } 
    \label{fig:pp-plot}
\end{figure} 
\begin{equation}
\epsilon = \frac{\left(\sum_i^N w_i\right)^2}{N \sum_i^N w_i^2} \in (0,100]\%\,,
\end{equation}
which depends on the distribution of the importance weights $w_i$ associated with the $N$ samples $\theta_i$ drawn from the flow.
The importance weights are defined as
\begin{equation}
\label{eq:importance_weights}
w_i = \frac{\mathcal{L}(d|\theta_i)\,\pi(\theta_i)}{q(\theta_i)}\,,
\end{equation}
where $\mathcal{L}(d|\theta)$ denotes the likelihood of the data $d$, $\pi(\theta)$ the prior distribution, and $q(\theta)$ the proposal distribution defined by the flow. 
The corresponding effective sample size is then given by $N_{\mathrm{eff}} = \epsilon  \times N$.

We assess the statistical calibration of the model using an ensemble of 1000 simulated signals drawn from the prior and injected into stationary Gaussian noise.
The results are summarized in the $p$--$p$ plot shown in Fig.~\ref{fig:pp-plot}, which illustrates the percentiles of the cumulative distribution function (CDF) of the marginal posterior at which the injected parameter values are recovered.
For all parameters, the percentile distributions are consistent with uniform distributions, indicating well-calibrated results.
This is further supported by a Kolmogorov-Smirnov (KS) test, which does not reject the null hypothesis of uniformity at the 95\% significance level and yields a combined $p$--value of 0.27 using Fisher's method~\cite{fisher1970statistical}. 
We note that this calibration study requires a full parameter-estimation run for each simulated signal; 
an equivalent analysis with traditional likelihood-based inference pipelines would therefore be computationally prohibitive.
Furthermore, across the analyzed events, we obtain a median sampling efficiency of $\epsilon\sim22\%$, consistent with previous application of \textsc{Dingo-IS} to LIGO-Virgo-KAGRA observations~\cite{2023PhRvL.130q1403D}. 

\renewcommand{\arraystretch}{1.3}
\begin{table}[t]
\centering
\caption{Injected parameter values for the three representative sources at low, moderate, and high SNR used in this work.}
\label{tab:injections}
\begin{tabular}{l@{\hskip 0.2cm}||@{\hskip 0.2cm}r@{\hskip 0.2cm}|@{\hskip 0.2cm}r@{\hskip 0.2cm}|@{\hskip 0.2cm}r}
\multirow{2}{*}{Parameter} 
    & Low SNR 
    & Moderate SNR 
    & High SNR \\ 
    & ${\rm SNR} = 87$
    & ${\rm SNR} = 500$ 
    & ${\rm SNR} = 1000$ \\[2pt]
\hline
&&&\\[-11pt]
$\mathcal{M}_c\,[10^{7} M_\odot]$ 
& \LowInjMc 
& \ModInjMc 
& \HighInjMc \\

$q$ 
& \LowInjq 
& \ModInjq 
& \HighInjq \\

$\chi_1$ 
& \LowInjChiOne 
& \ModInjChiOne 
& \HighInjChiOne \\

$\chi_2$ 
& \LowInjChiTwo 
& \ModInjChiTwo 
& \HighInjChiTwo \\

$d_L\,[\mathrm{Gpc}]$ 
& \LowInjdL 
& \ModInjdL 
& \HighInjdL \\

$\theta_{JN}\,[\mathrm{rad}]$ 
& \LowInjThetaJN 
& \ModInjThetaJN 
& \HighInjThetaJN \\

$\theta_S\,[\mathrm{rad}]$ 
& \LowInjThetaS 
& \ModInjThetaS 
& \HighInjThetaS \\

$\phi_S\,[\mathrm{rad}]$ 
& \LowInjPhiS 
& \ModInjPhiS 
& \HighInjPhiS \\

$t_c\,[\mathrm{s}]$ 
& \LowInjTc 
& \ModInjTc 
& \HighInjTc \\

$\psi\,[\mathrm{rad}]$ 
& \LowInjPsi 
& \ModInjPsi 
& \HighInjPsi \\

$\phi_c\,[\mathrm{rad}]$ 
& \LowInjPhiC 
& \ModInjPhiC 
& \HighInjPhiC \\
\end{tabular}
\end{table}
\renewcommand{\arraystretch}{1}
\subsection{Benchmarking}
\label{bench}
We benchmark \textsc{Dingo} against posterior samples obtained with the nested sampling algorithm~\cite{Skilling:2006gxv} as implemented in \textsc{Nessai}~\cite{2021PhRvD.103j3006W}.
To ensure a consistent comparison, both methods employ the same waveform model, prior distributions, analysis settings, and noise PSD described in Sec.~\ref{methods}.
\begin{figure*}[t]
    \centering
    \includegraphics[width=\textwidth, keepaspectratio]{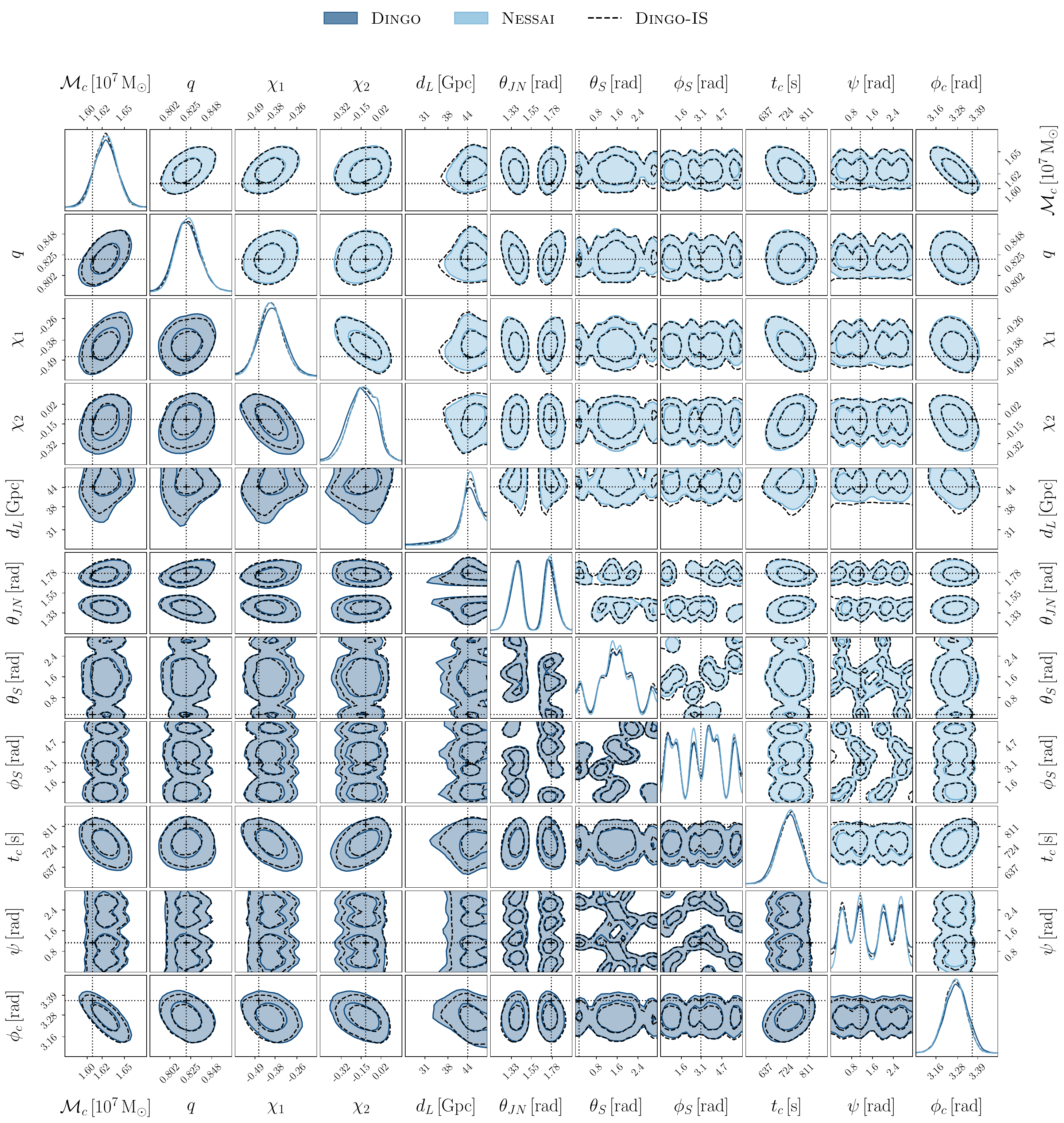}
    \caption{Posterior distributions for our representative source with low SNR ($\sim 87$) obtained with different sampling methods.
    The lower left triangle compares results from \textsc{Dingo} (dark-blue filled contours) and \textsc{Dingo-IS} (dashed black contours), while
    the upper right triangle compares \textsc{Nessai} (light-blue filled contours) and \textsc{Dingo-IS}.
    Contours indicate the 50\% and 90\% credible regions.
    The true injected values are indicated by dotted lines.}
    \label{fig:corner_low}
\end{figure*} 
We consider three representative signal injections spanning the targeted SNR range, corresponding to low, moderate, and high SNRs of approximately 87, 500, and 1000, respectively.
The injected parameters are reported in Table~\ref{tab:injections}.
\begin{figure*}[t]
    \centering
    \includegraphics[width=\textwidth, keepaspectratio]{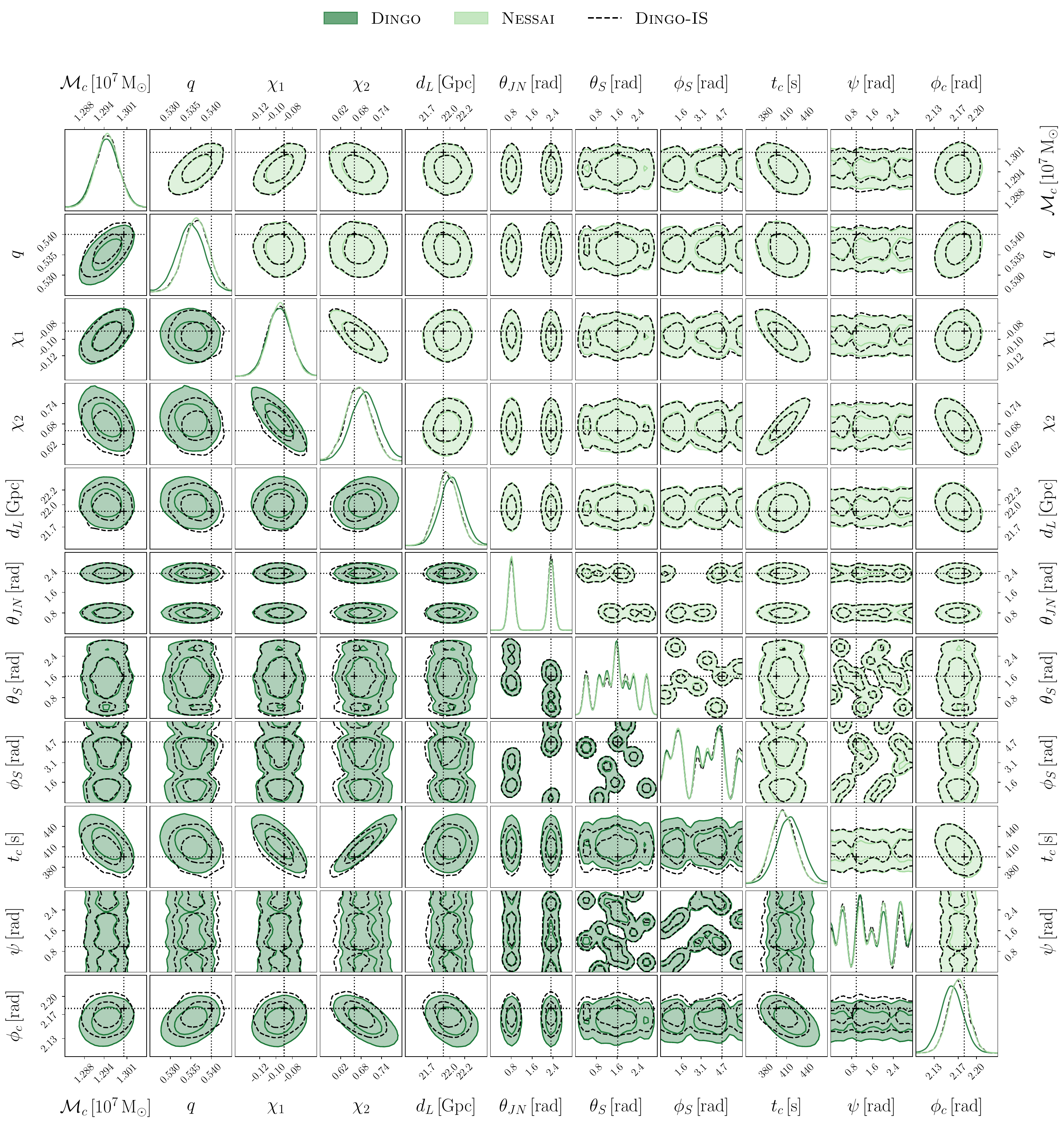}
    \caption{Posterior distributions for our representative source with moderate SNR ($\sim 500$) obtained with different sampling methods.
    The lower left triangle compares results from \textsc{Dingo} (dark-green filled contours) and \textsc{Dingo-IS} (dashed black contours), while
    the upper right triangle compares \textsc{Nessai} (light-green filled contours) and \textsc{Dingo-IS}.
    Contours indicate the 50\% and 90\% credible regions.
    The true injected values are indicated by dotted lines.} 
    \label{fig:corner_moderate}
\end{figure*} 
\begin{figure*}[t]
    \centering
    \includegraphics[width=\textwidth, keepaspectratio]{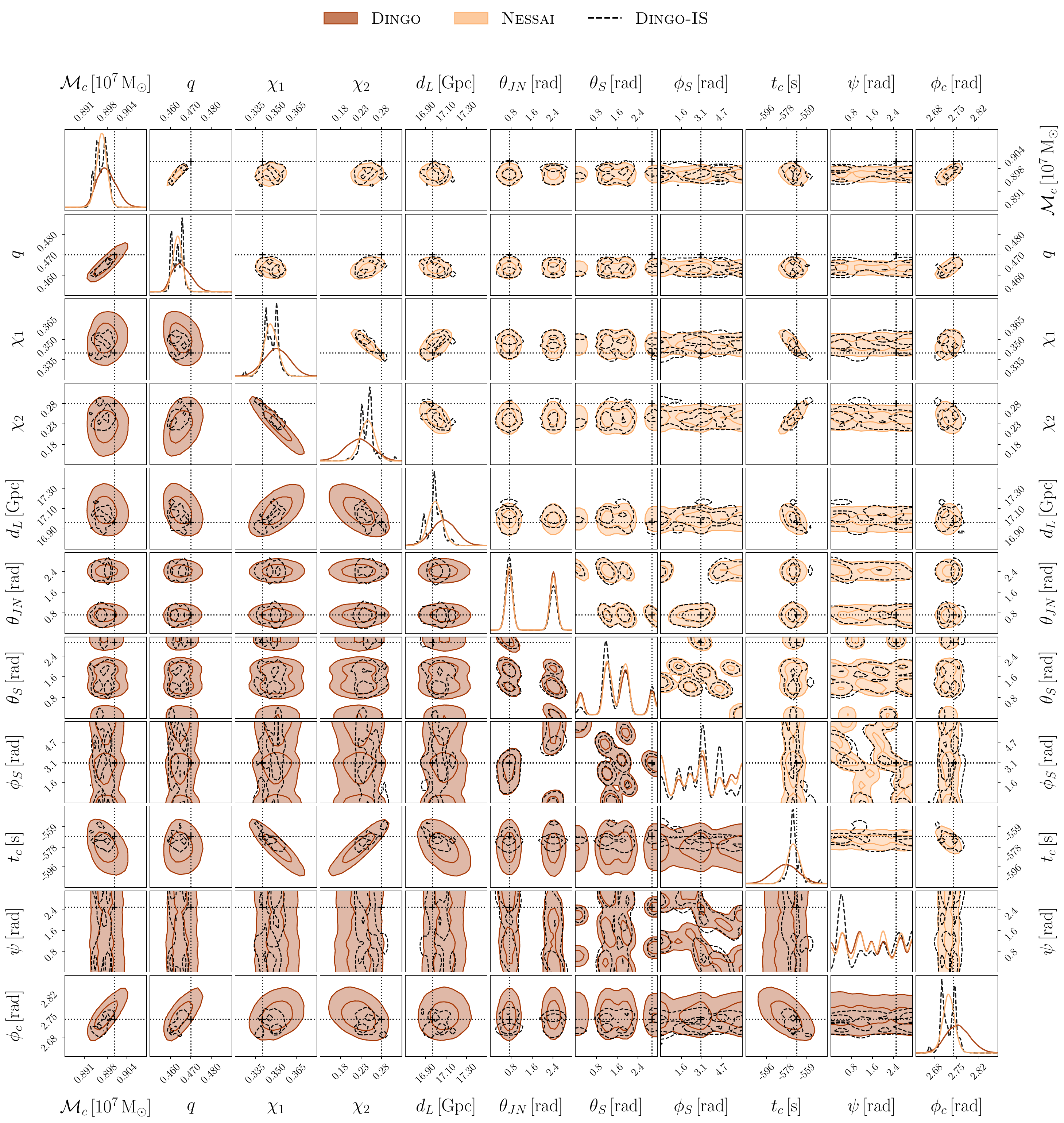}
    \caption{Posterior distributions for our representative source with high SNR ($\sim 1000$) obtained with different sampling methods.
    The lower left triangle compares results from \textsc{Dingo} (dark-orange filled contours) and \textsc{Dingo-IS} (dashed black contours), while
    the upper right triangle compares \textsc{Nessai} (light-orange filled contours) and \textsc{Dingo-IS}.
    Contours indicate the 50\% and 90\% credible regions.
    The true injected values are indicated by solid dotted lines.} 
    \label{fig:corner_high}
\end{figure*} 
For each injection, we draw $2\times10^{4}$ samples from the flow posterior.
Initial sampling requires less than two seconds on a single 40 GB NVIDIA A100 GPU, while IS adds an additional ten to fifteen seconds on 20 CPU cores, resulting in a full-parameter inference time well below one minute per event.
By comparison, the corresponding \textsc{Nessai} runs were performed using 4000 live points and without acceleration or optimization techniques.
The runtime ranged from roughly ten days for the low-SNR case to about forty days for the high-SNR scenario.
These values likely represent a pessimistic estimate, as GPU-accelerated analyses combined with heterodyned likelihood techniques (see e.g. Ref.~\cite{2022PhRvD.105d4055K}) have been shown to reduce the computational cost of stochastic sampling methods to tens of minutes. 
Nevertheless, the inference approach presented in this work remains significantly less computationally demanding.

In Fig.~\ref{fig:corner_low}, we show the posterior distributions for the low-SNR injection, comparing samples obtained from the \aur{flow-based}{raw} \textsc{Dingo} output, after IS correction (\textsc{Dingo-IS}), and from the reference \textsc{Nessai} analysis.
\textsc{Dingo} and \textsc{Nessai} posteriors are in very close agreement, with the corresponding \textsc{Dingo-IS} results fully overlapping both. 
This indicates that IS introduces only negligible corrections in this regime, a conclusion further supported by the recovered sample efficiency, $\epsilon = 50\%$, which corresponds to an effective sample size of $N_{\mathrm{eff}} = 10^4$.
Beyond the overall agreement, \textsc{Dingo} successfully captures the non-trivial structure of the posterior, including correlations between luminosity distance $d_L$, inclination $\theta_{JN}$, and sky location ($\theta_S, \phi_S$).
In particular, the characteristic multimodal sky structure induced by the low-frequency response of LISA~\cite{2021PhRvD.103h3011M} is consistently recovered.

Figures~\ref{fig:corner_moderate} and~\ref{fig:corner_high} show the posterior distributions for the moderate- and high-SNR sources, respectively.
At moderate SNR, the \aur{flow-based}{raw} \textsc{Dingo} posteriors display small but noticeable deviations from the \textsc{Nessai} reference.
These residual discrepancies are effectively compensated by IS, bringing the resulting \textsc{Dingo-IS} posteriors into close agreement with \textsc{Nessai}. 
In this case, we report a sample efficiency of $\epsilon= 12\%$ and an effective sample size of $N_{\rm eff}=2.4 \times10^3$. 

For the high-SNR injection, the \textsc{Dingo} posteriors exhibit broader support than the \textsc{Nessai} reference, resulting in limited overlap between the two distributions. 
In this case, the learned proposal distribution assigns non-negligible probability mass to regions of low true posterior density, leading to highly variable importance weights (Eq.~\ref{eq:importance_weights}) and a sampling efficiency of $\epsilon = 0.05\%$. 
The resulting posterior estimate obtained with \textsc{Dingo-IS} remains dominated by statistical fluctuations, as illustrated by the one- and two-dimensional marginal distributions in Fig.~\ref{fig:corner_high}, due to the extremely small effective sample size ($N_{\mathrm{eff}} = 10 \ll N$).
Increasing the number of flow samples might improve the stability and smoothness of the reweighted posterior.
Despite this, it is worth noting that the \aur{flow-based}{raw} \textsc{Dingo} posterior correctly localizes the source within a restricted region of the parameter space, reducing the original prior volume by approximately a factor of $10^{13}$ at the 90\% credible level.
The corresponding one- and two-dimensional marginal distributions also provide good coverage of the target distribution, offering a reliable starting point for subsequent refinement with conventional stochastic sampling methods.
The sky location and polarization parameters remain weakly constrained owing to intrinsic degeneracies in the adopted LISA response. 
As in the low- and moderate-SNR injections, the posteriors are highly multimodal and extend over most of the prior range.
Nonetheless, the model captures the characteristic sky pattern expected in the low-frequency regime.

\begin{figure*}[t]
    \centering
    \includegraphics[width=\textwidth, keepaspectratio]{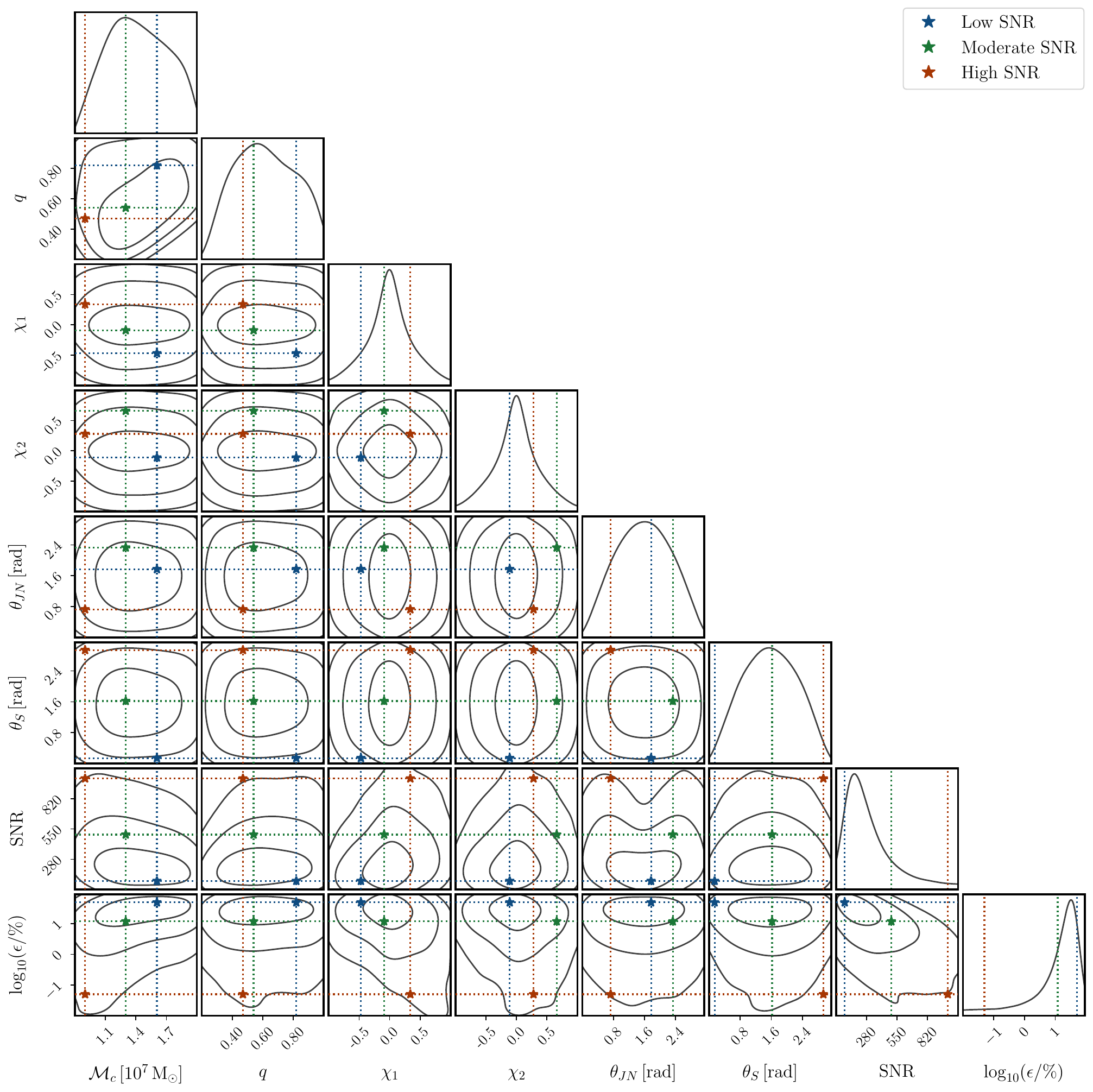}
    \caption{Prior distributions used to construct the target training set.
    Injected values for the low- (blue), moderate- (green), and high-SNR (orange) cases are shown as dotted lines, with colored stars marking their locations.
    Contours contain 50\%, 90\%, and 99\% of the estimated prior density.
    Of note, the bottom right section shows the joint and marginal distributions of the SNR and sample efficiency $\epsilon$ across the ensemble of simulated injections.} 
    \label{fig:prior_coverage}
\end{figure*}

To shed light on the origin of the low sampling efficiency observed for the high-SNR injection, we investigate how it varies across the source parameter space.
This is illustrated in Fig.~\ref{fig:prior_coverage}, which shows the prior distribution used to construct the training dataset, together with the injected parameter values of the three representative BH binaries described above. 
The visualization is restricted to parameters with non-uniform priors, since only these imprint a non-trivial structure on the sampling density.
We draw $2\times 10^4$ sources from this prior and evaluate their posterior distributions using our trained model.
The resulting SNR and sampling efficiency $\epsilon$ are reported in Fig.~\ref{fig:prior_coverage} as a function of the source parameters.
Crucially, we observe a strong anti-correlation between SNR and sampling efficiency.
This behaviour can be understood as a consequence of the mass-covering property of the training objective used in neural posterior estimation.
Specifically, the loss penalizes the density estimator when it assigns insufficient probability mass to regions supported by the true posterior, thereby favoring approximations that cover its full support.
In the presence of residual modeling inaccuracies, this property typically results in overly diffuse posterior distributions.
While avoiding posterior undercoverage is generally desirable in statistical inference, this tendency becomes particularly problematic in the high-SNR regime, where the true posterior is sharply concentrated.
In this situation, even a mild overestimation of the posterior support causes a substantial fraction of samples to fall in low-likelihood regions, resulting in highly variable importance weights and a corresponding reduction in the effective sample size.

An additional factor contributing to the observed reduction in sampling efficiency is the coverage of the training prior.
As illustrated in Fig.~\ref{fig:prior_coverage}, the high-SNR source lies in a sparsely represented region of this distribution, with the injected chirp mass $\mathcal{M}_c = 9 \times 10^6\,\mathrm{M_\odot}$ falling below the $3.5^{\mathrm{th}}$ percentile. 
Consequently, low chirp-mass systems are rarely encountered during training and therefore contribute less to the optimization objective, leading to a poorer approximation of the posterior and a corresponding decrease in sampling efficiency.
A similar effect is observed for the primary spin.
Although the injected value, $\chi_1 = 0.34$, lies well within the allowed range, the peaked structure of its prior places it above the $85^{\mathrm{th}}$ percentile of the training prior.

To explore this dependence, we vary the chirp mass and primary spin values while keeping all other parameters fixed to the reference values listed in Tab.~\ref{tab:injections}.
The luminosity distance $d_L$ and source inclination $\theta_{JN}$ are adjusted to maintain a constant SNR.
Increasing the chirp mass to $1.3 \times 10^7\,\mathrm{M_\odot}$ shifts the system toward a more densely sampled region of the training prior and raises the sampling efficiency to $1.5\%$. 
A similar improvement is obtained by setting the primary spin to $\chi_1 = 0$, which increases the efficiency to $0.5\%$.
By contrast, varying the sky colatitude $\theta_S$ has little effect on the sampling efficiency;
even when moved toward more densely sampled regions of the training prior, the efficiency remains close to the reference value of $\epsilon = 0.05\%$.
These results indicate that underrepresentation in the training prior does not affect all parameters equally. 
Waveforms with low chirp masses and high primary spins exhibit more oscillatory structure and are therefore more difficult to model accurately, making sparse training coverage in these regions particularly detrimental to the performance of the density estimator.

We emphasize that limitations in prior coverage are not exclusive to high-SNR sources.
Indeed, among the 1000 injections used to construct the $p$--$p$ plot presented in Sec.~\ref{model validation}, we identify a small subset of sources ($\sim$ 10) with SNR $\in[250,500]$ and sample efficiency below 1\%.
These systems typically have parameters located near the boundaries of the training prior, most commonly featuring either high primary spins above the $95^{\mathrm{th}}$ percentile or low chirp masses below the $5^{\mathrm{th}}$ percentile. 
Two sources differ from this pattern: one has a mass ratio above the $95^{\mathrm{th}}$ percentile together with an anti-aligned primary spin below the $5^{\mathrm{th}}$ percentile, while the other exhibits strongly anti-aligned spins, with $\chi_1$ and $\chi_2$ below the $0.1^{\mathrm{th}}$ and $4^{\mathrm{th}}$ percentiles, respectively.

Overall, our results show that sampling efficiency systematically decreases at high SNR.
Extending the training does not significantly improve the sampling efficiency, suggesting that insufficient optimization is unlikely to be the primary source of the observed mismatch. 
Further improvements may instead require increased model capacity or more flexible conditional density parameterizations.
In addition, the sensitivity of the sampling efficiency to peripheral regions of parameter space indicates that the adopted training prior plays a non-negligible role. 
This dependence could be mitigated by broadening the prior support or by adopting a distribution that provides more uniform coverage of the most informative parameters.
A systematic exploration of the full LISA mass spectrum and its implications for prior coverage is left for future work.

Finally, as done in previous works~\cite{2021PhRvL.127x1103D,2023PhRvL.130q1403D,2025PhRvD.112j3015S}, we use the Jensen-Shannon (JS) divergence~\cite{61115} to quantitatively compare the marginal posterior distributions obtained with \textsc{Dingo} and \textsc{Nessai} across 100 independent injections for each representative source considered in this study.
As discussed above, posterior recovery in the high-SNR case already shows qualitative differences, and we therefore leave a quantitative assessment of this regime to future work.
The injections are generated using the same noise realization for each representative source, thereby isolating sampling errors. 
The JS divergence ranges from 0 bit to 1 bit, corresponding to identical and maximally distinct distributions, respectively.
Across all parameters, we find median divergencies of $D^{L}_{JS}\in [6.3\times10^{-4}, 2\times10^{-2}]$ for the low-SNR injections and $D^{M}_{JS}\in[1.2\times10^{-4}, 0.1]$ for the moderate-SNR case.
Applying IS to the \textsc{Dingo} posteriors leads to a systematic reduction of the divergence in both regimes, yielding $D^{L}_{JS}\in [1.9\times10^{-4}, 1\times10^{-2}]$ and $D^{M}_{JS}\in[1.7\times10^{-4}, 7.3\times10^{-4}]$, respectively.
For a reference scale, let us consider the JS divergence between two univariate Gaussian distributions with mean $\mu_{1,2}$ and standard deviation $\sigma_{1,2}$.
The largest median divergence observed in our analysis ($D^{M}_{JS}=0.1$, corresponding to the phase parameter of the moderate-SNR source) is comparable either to a mean offset of $|\mu_1-\mu_2|\simeq0.8\,\sigma$ for a case where $\sigma=\sigma_1=\sigma_2$, or to a standard-deviation ratio $\sigma_1/\sigma_2\simeq1.75$ for a case with $\mu_1=\mu_2$.
For the low-SNR case, the largest median divergence ($D^{M}_{JS}=2\times10^{-2}$) is obtained for the luminosity distance parameter.
When mapped onto the Gaussian reference, this corresponds either to a mean offset of $|\mu_1-\mu_2|\simeq0.3\,\sigma$ for equal variances, or to a standard-deviation ratio $\sigma_1/\sigma_2\simeq1.3$ for equal means.
After IS, a substantial improvement is observed, particularly in the moderate-SNR case, where the largest median JS divergence corresponds to $|\mu_1-\mu_2|\simeq0.06\,\sigma$ and  $\sigma_1/\sigma_2\simeq1.04$.
Simultaneous differences in mean and variances smaller than the values provided yield smaller JS divergences.

\subsection{Comparison with previous work}
\label{previous}
Previous works have applied SBI techniques to the analysis of massive BH binaries.
Although these studies primarily focused on lower-mass systems than those considered here, in the following we present a qualitative comparison of the underlying methodological approaches.

In contrast to the present work, Ref.~\cite{2025JCAP...04..022M} adopts a neural likelihood estimation strategy, training neural networks to construct a surrogate of the likelihood function.
This method achieves a substantial reduction in the number of waveform evaluations relative to traditional stochastic samplers, while yielding posterior distributions that are qualitatively consistent with standard likelihood-based results.
However, a comprehensive quantitative validation against established methods is still lacking, preventing a robust assessment of its overall performance.
Inference also relies on a dedicated training process for each individual source, with reported computational costs of the order of days, thereby limiting the applicability of this method to time-critical analyses.

By comparison, Ref.~\cite{2024MLS&T...5d5040L} adopts a neural posterior-estimation framework more closely aligned with the approach presented in this work, based on continuous normalizing flows~\cite{2018arXiv180607366C} trained using the flow matching algorithm~\cite{2022arXiv221002747L}.
The inferred posterior distributions show qualitative agreement with those obtained using nested sampling; 
however, systematic deviations persist across multiple parameters, indicating that further methodological refinement may be required to achieve inference accuracy comparable to that achieved in this work.
In particular, model performance is evaluated under different noise settings: Galactic foreground noise is included in the test set but not during training, while the nested-sampling analysis explicitly accounts for it.
Consequently, the two approaches are not assessed under matched conditions, and direct comparisons should be treated with caution.
From a computational perspective, their network features approximately 200 million learnable parameters and trains in about two days, whereas the discrete normalizing-flow model considered here has roughly 350 million parameters and requires approximately 10 days of training.
However, per-event inference typically requires several minutes to generate $\sim1.4\times10^4$ posterior samples, compared to a few seconds in this work. 

\section{Outlook}
\label{appl and challenges}
The successful demonstration of \textsc{Dingo} as an accurate and efficient framework for full parameter estimation of massive BH binary signals enables a broad range of applications for LISA data analysis.
Among these, low-latency pipelines are particularly compelling, as they are crucial for enabling coordinated observations of the diverse electromagnetic signatures expected to accompany massive BH binaries~\cite{2022LRR....25....3B,2023arXiv231016896D,2012AdAst2012E...3D,2019NewAR..8601525D}.
The sky-localization uncertainties anticipated for LISA observations of such sources are expected to encompass numerous potential host galaxies, posing a significant challenge for unambiguous host identification~\cite{2023MNRAS.519.5962L}.
Rapid inference of additional source properties ---most notably the mass and luminosity distance--- is therefore essential to provide independent constraints on the host environment and to efficiently reduce the potential host-galaxy candidates.
In this context, developing a \textsc{Dingo} network trained on pre-merger signals (see Ref.~\cite{2025Natur.639...49D} for an application to binary neutron-star systems) and integrating it within low-latency search pipelines would be a promising step forward toward LISA parameter estimation in realistic early-warning scenarios.

Beyond low-latency applications, the fast inference capabilities of \textsc{Dingo} can be leveraged within the LISA global fit to substantially reduce the computational cost associated with massive BH binary analyses. 
More specifically, this can be achieved by integrating the proposed SBI approach within a blocked Gibbs framework~\cite{ritter1992facilitating,muller1991genetic,gilks1992adaptive}. 
In this hybrid scheme, \textsc{Dingo} could generate samples for individual sources at each global-fit iteration, either replacing Markov Chain Monte Carlo sampling~\cite{2010CAMCS...5...65G} when the learned posterior approximation is sufficiently accurate, or acting as a proposal distribution to accelerate convergence otherwise.
This will facilitate the adoption of more accurate waveform models, whose computational expense would otherwise be prohibitive for fully stochastic inference frameworks.
Extending this approach to self-consistent global-fit implementations would require relaxing the fixed-PSD assumption adopted in this work and training the network on a representative ensemble of instrumental noise and Galactic foreground realizations, thereby accounting for event-to-event variations in the noise model. 
In the signal-dominated regime expected for LISA, the noise PSD cannot be estimated independently of the GW signals and must instead be inferred as part of the global fit.
In this context the PSD estimated at a previous stage of the analysis could be used to condition \textsc{Dingo} at inference time, ensuring consistency between the source parameter estimation and the evolving noise model.

A further challenge common to both low-latency and global-fit pipelines is the presence of non-stationary features in the data, such as glitches and data gaps.
In global-fit analyses, glitches are either incorporated directly into the data model and inferred jointly with the astrophysical signals~\cite{2023PhRvD.108l3029S,2025PhRvD.112f3041M,2019PhRvD..99b4019R}, or mitigated by effectively treating them as data gaps through direct masking~\cite{2025CQGra..42f5018C}.
In parallel, machine-learning–based approaches have been developed for the rapid identification and mitigation of noise transients in low-latency settings~\cite{2024PhRvD.109h3027H}.
However, these methods are not designed to deliver full end-to-end parameter estimation.
Data gaps have also been addressed through a variety of techniques, including apodization~\cite{2021PhRvD.104d4035D}, inpainting~\cite{2022MNRAS.509.5902B}, data augmentation~\cite{2019PhRvD.100b2003B,2025PhRvD.111b4067M}, and non-diagonal likelihood formulations~\cite{2025PhRvD.111l4053B}.
These approaches often come at the cost of additional noise-model approximations, potential biases introduced by gap-filling procedures, or significant computational overhead.
More recently, Ref.~\cite{2025arXiv250905479P} proposed an efficient wavelet-domain augmentation scheme;
however, its applicability relies on the assumption of locally stationary noise.

Finally, future work will explore the performance of \textsc{Dingo} on precessing and potentially eccentric signals, as well as its extension to lower-mass systems emitting GW signals that typically last weeks to months.
In this regime, the breakdown of the low-frequency approximation requires the full time- and frequency-dependent LISA response.
These effects substantially increase the effective input dimensionality and the computational cost of training, motivating the investigation of data compression techniques, such as multibanding \cite{2025Natur.639...49D}, alongside the development of more flexible model architectures.\\

Taken together, these considerations highlight the need for fast, preliminary source-characterization methods that remain effective in the presence of realistic, time-varying noise and physics-rich waveform models. 
SBI is the obvious answer here, as noise artifacts can be included in the training data without the need to specify a likelihood function (note, however, that IS ---used both in this paper and in previous {\sc Dingo} work--- \emph{does} require evaluating the likelihood).
In this context, Ref.~\cite{2025arXiv251218290M} has demonstrated that SBI provides a flexible and efficient framework capable of handling  data gaps.
This capability is particularly promising for time-critical analyses and may also facilitate the initialization of global-fit pipelines by improving convergence and robustness in complex noise environments.
The recent transformer-based extension of \textsc{Dingo}~\cite{2025arXiv251202968K}, designed to accommodate variations in data-analysis settings, represents a natural next step that builds on the approach presented in this work toward addressing key challenges for LISA, such as data gaps and source overlap (see also Ref.~\cite{2025CQGra..42r5012P,2025arXiv251221390B,2025arXiv251123228H}). 

$\,$
\begin{acknowledgments}
We thank Maximilian Dax, Annalena Kofler, Federico De Santi, Philippa Cole, Filippo Santoliquido, Rahul Srinivasan, and Golam Shaifullah for discussions.
A.S., D.G., R.B. and R.T. are supported by MUR Grant ``Progetto Dipartimenti di Eccellenza 2023-2027'' (BiCoQ),
and the ICSC National Research Centre funded by NextGenerationEU. 
A.S., D.G., and R.T. are supported by 
ERC Starting Grant No.~945155--GWmining, 
Cariplo Foundation Grant No.~2021-0555, 
and
Italian-French University (UIF/UFI) Grant No.~2025-C3-386.
D.G. is supported by MSCA Fellowship No.~101149270--ProtoBH and MUR Young Researchers Grant No. SOE2024-0000125.
S.R.G. is supported by a UKRI Future Leaders Fellowship (grant number MR/Y018060/1).
R.B. is supported by the Italian Space Agency grant ``Phase
B2/C activity for LISA mission'', Agreement n.2024-NAZ-0102/PE.
M.P. is supported by NSF Grants AST-2407453 and
PHY-2512902.
Computational work was performed on the {Saraswati} and {Hypatia} clusters at AEI (Potsdam). 

\end{acknowledgments}

\appendix

\bibliographystyle{apsrev4-2} 
\bibliography{dingolisa}

\end{document}